\documentclass[runningheads,citeauthoryear]{apinv}
\usepackage{epsfig,graphics}
\usepackage{cite}
\usepackage{marvosym}
\usepackage[T2A]{fontenc}
\usepackage[utf8]{inputenc}
\usepackage{float}
\usepackage{tabu}

\begin{document}

\title{Rotation period determination for asteroid 9021 Fagus\thanks{Based on data collected with 2m RCC telescope at Rozhen National Astronomical Observatory.}}
											
\titlerunning{Rotation period of 9021 Fagus ...}
\author{G. Apostolovska\inst{1}, A. Kostov\inst{2}, Z. Donchev\inst{2} and E. Vchkova Bebekovska\inst{1}}
\authorrunning{G. Apostolovska et al.}
\tocauthor{G. Apostolovska, A. Kostov, Z. Donchev and E. Vchkova Bebekovska}
\institute{
\inst{1}Institute of Physics, Faculty of Science, Ss. Cyril and Methodius University, Skopje, Republic of Macedonia, Arhimedova 3, 1000 Skopje, Republic of Macedonia
\newline
\inst{2}Institute of Astronomy and NAO, Bulgarian Academy of Sciences, Tsarigradsko Chaussee Blvd. 72, 1784, Sofia, Bulgaria
\newline
	\email{gordanaaspostolovska@gmail.com}
}

\papertype{Submitted on 15.05.2017. Accepted on 11.06.2017.}	
\maketitle

\begin{abstract}Lightcurve analysis of the asteroid 9021 Fagus observed at the Bulgarian National Astronomical Observatory Rozhen during two apparitions in 2013 and 2017 are presented. This asteroid was observed in 2013 accidentally in the field of view in which our long-term target asteroid 901 Brunsia was positioned.  A search of the Asteroid Lightcurve Database (Warner et al. 2009) did not find any previously reported results for the rotation period of 9021 Fagus. Unfortunately two nights observations in 2013 showed that the rotation period could be only approximately assumed. The observations during two neighbouring nights in 2017 March 20 and 21  revealed the period of  $5.065 \pm 0.002$ h with amplitude of $0.73\pm0.02$ mag. Obtained lightcurves in combination with other observational techniques, as well as with data gathered from future space mission will contribute to the enlargement of the database for rotational characteristics of the asteroids.
\end{abstract}


\section{Introduction}

Astrometric and photometric investigations of asteroids have long tradition at the Bulgarian National Astronomical Observatory (BNAO) Rozhen.  In the frame of the astrometric program many of new asteroids were discovered and some of them were named by the discoverers (Ivanova et al. 2002). The importance of the photometric investigations of asteroids is significant from two main aspects. The scientific one is connected with the fact that asteroids are  remnants of the original material from which our solar system bodies formed and  knowing the physical characteristics, the distribution and the evolution of asteroids is crucial for the understanding of the Solar system creation.  Beside the scientific aspect, the investigations of the asteroids are important for human civilization for many reasons. One of this is discovering of dangerous asteroids and protection of the planet and life on it. Some of asteroids being considered as sources of rare minerals and metals are interesting from an economical viewpoint. In the last years some Near Earth asteroids are considered as potential target for manned mission as training for future manned mission to the planet Mars.  For all of those reasons in December 2016 United Nations in its resolution declares that 30 June will be International Asteroid day and 2017 will be first year when this event will be officially celebrated around the world\footnote{https://asteroidday.org/}. 

\section{Observations and data reduction}

The photometric observations of 9021 Fagus were performed at the BNAO Rozhen. In 2013 they were made by $50/70$ Schmidt telescope with an FLI PL 16803 CCD  ($4096\times4096$ pixels, pixel size $9\mu m$). For the photometric observations in 2017 we used 2-channel focal reducer – Rozhen (FoReRo2) of the 2m Ritchey-Chr\'{e}tien-Coude (RCC) telescope with VersArray 1300B  CCD ($1340\times1300$ pixels, pixel size $20\mu m$).

For observations obtained with the Schmidt telescope we performed light images reduction using dark subtraction and flat-field correction. For 2-m telescope observations we made light images reduction including only bias and flat-field correction. Using the software program CCDPHOT by Buie (1998) we performed aperture photometry of the asteroid and the comparison stars. For lightcurve analysis, we used software package MPO Canopus v10.7.7.0\footnote{MPO Canopus Software.http://www.MinorPlanetObserver.com} (Warner  2016), that produces reduced composite lightcurves from several nights, calculates rotational period and can make Fourier analysis and estimates of the amplitude of the lightcurve.

\section{Results}

Asteroid 9021 Fagus belongs to the middle part of the main belt and it orbits around the Sun for the  period of 4.14  years. According to the WISE (Wide-field Infrared Survey Explorer) and NEOWISE (Masiero et al. 2011) it has a diameter of 13.096 km and albedo 0.124 (JPL Small-Body Database Browser\footnote{https://ssd.jpl.nasa.gov/sbdb.cgi\#top}). 9021 Fagus was discovered by Belgian astronomer E. W. Elst on February 14, 1988 at the La Silla observatory. E.W. Elst is ranking by the Minor Planet Center among the top 10 discoverers of 3866 minor planets made between 1986 and 2009 at ESO La Silla Observatory and at the BNAO Rozhen\footnote{https://en.wikipedia.org/wiki/Eric\_Walter\_Elst}. This asteroid was named for the plant genus Fagus, which includes trees commonly known as beech or beech family (Fagaceae).

During two nights in 2013 March 10 and 12 on Schmidt telescope we actually observed the asteroid 901 Brunsia which was our long-term target for shape modelling. The asteroid was observed both nights about four hours which covered more than one rotational period. It was about 15.4 mag and we used exposure time of 240s.  In the same images beside Brunsia, we noticed  very faint asteroid which was identified as 9021 Fagus and has no published rotation period. The used exposure time was very short for this asteroid, which was at about 16.95 mag  and was  faint target for the Schmidt telescope. But the very short exposure time did not make problems during lightcurve analysis because the weather conditions were  good and mainly because accidentally the asteroid had very high amplitude of its rotational curve. The opposition of 9021 Fagus in 2013 was on February 19 and we observed  the asteroid after its opposition at the solar phase angle increased from $7.7^{\circ}$ to  $8.5^{\circ}$. All images were taken through an R filter.

\begin{figure}[h]
\begin{center}
\hspace{-20 mm}
    \mbox{\epsfig{file=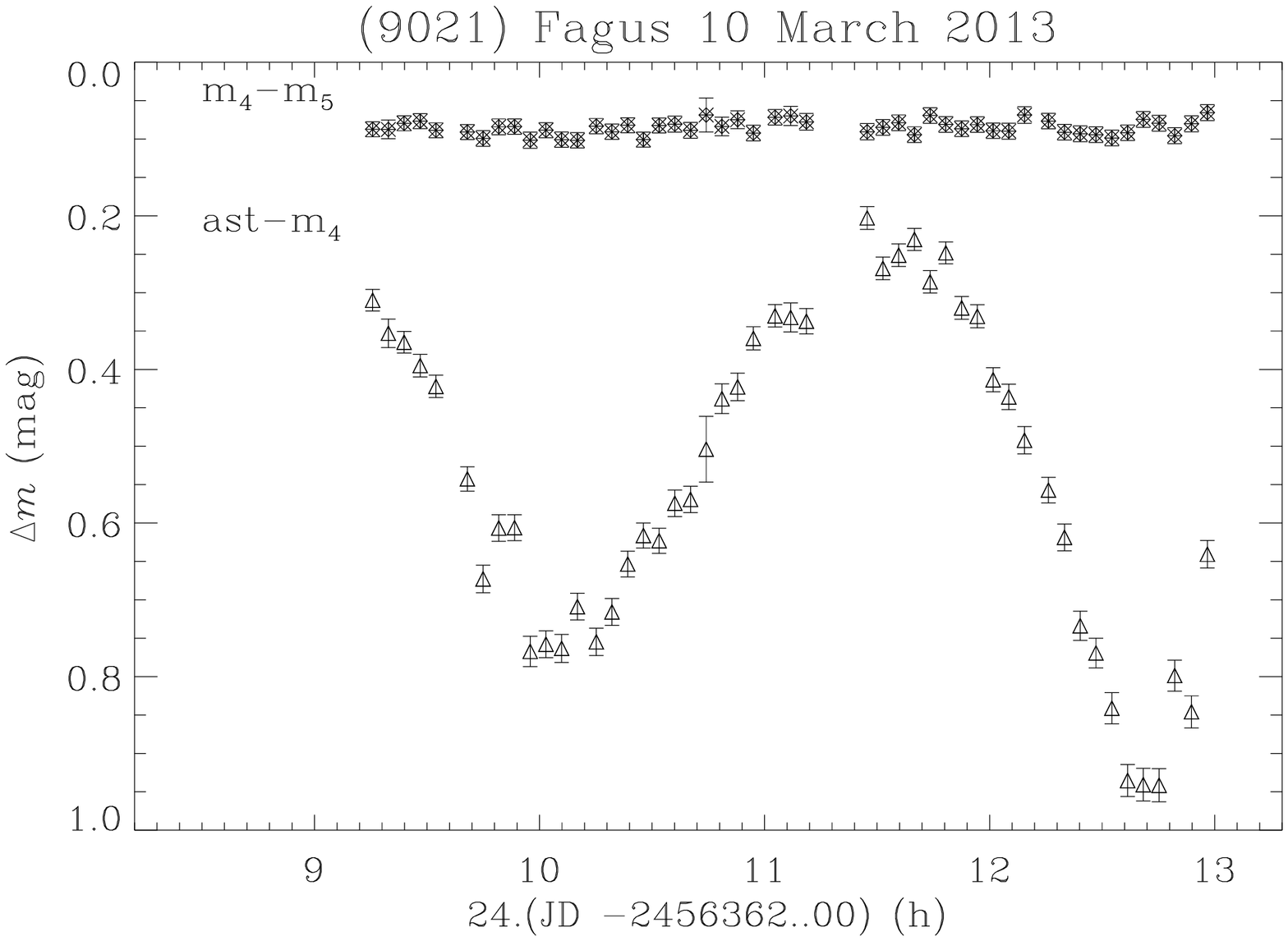, width=0.3\textwidth,angle=90}}   
    \hspace{10 mm}
    \mbox{\epsfig{file=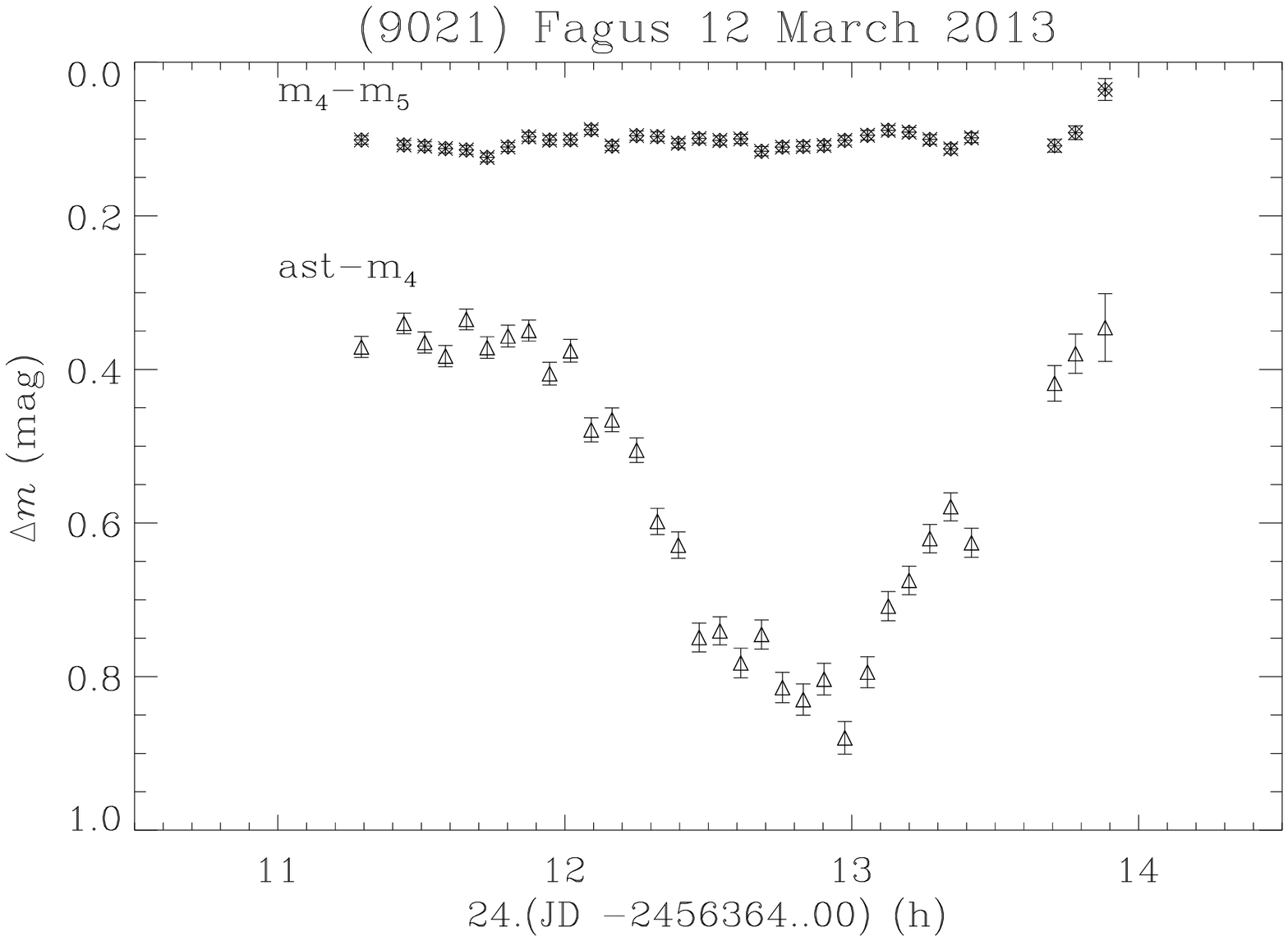, width=0.3\textwidth,angle=90}}
		\vspace{20 pt}
    \caption{Relative light curve of 9021 Fagus on 10 March 2013 (left) and 12 March 2013 (right).}
		\label{9021March2013LC}
\end{center}
\end{figure}
From the partial lightcurves presented in Fig.\ref{9021March2013LC} and assuming two almost symmetrical minima and maxima for the rotational lightcurve we could only give some preliminary result for the period of about 5.5 hours with error more than 1 hour.

\begin{figure}[h]
\begin{center}
    \mbox{\epsfig{file=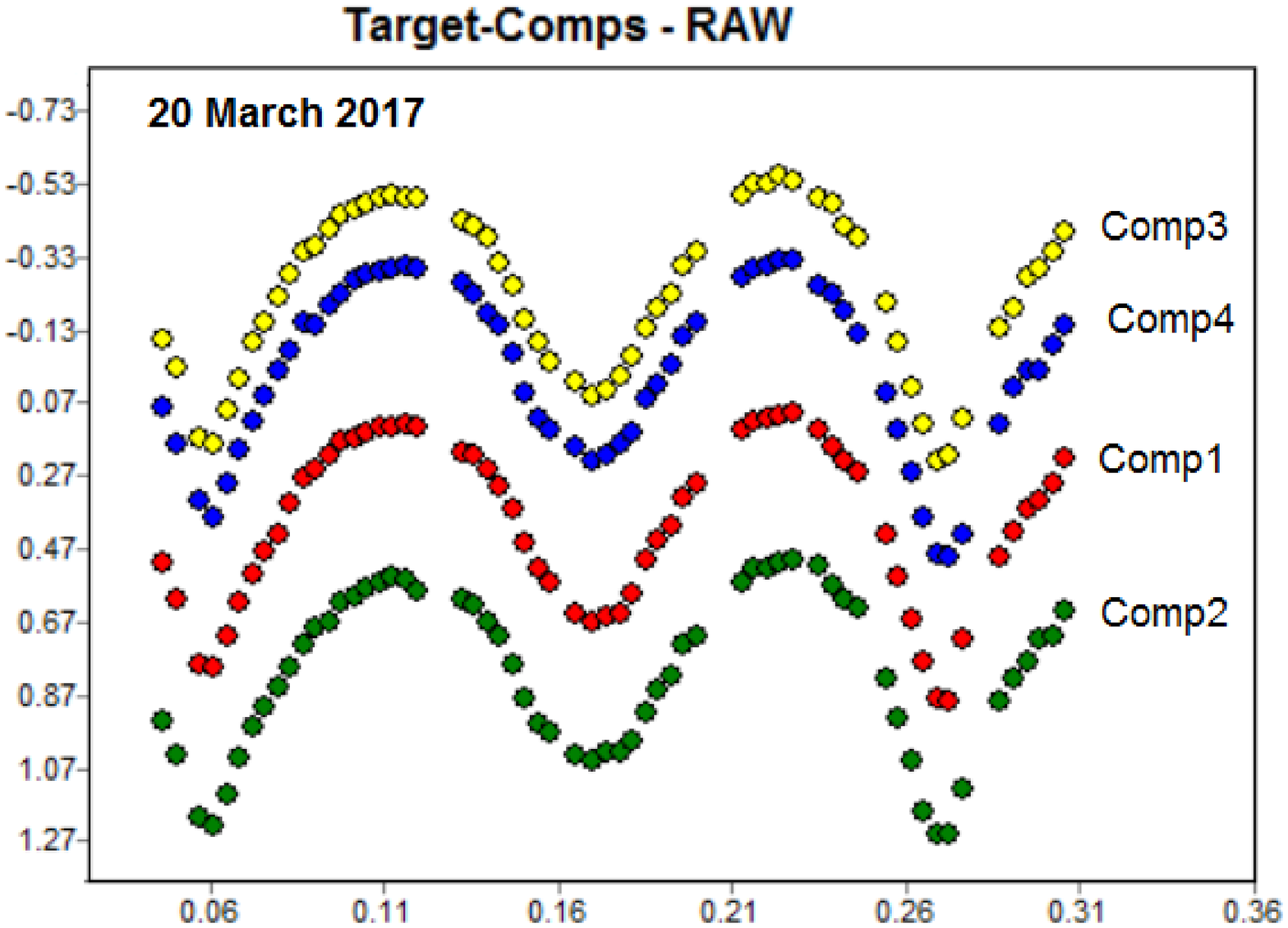, width=0.45\textwidth}}   
    \hspace{10 mm}
    \mbox{\epsfig{file=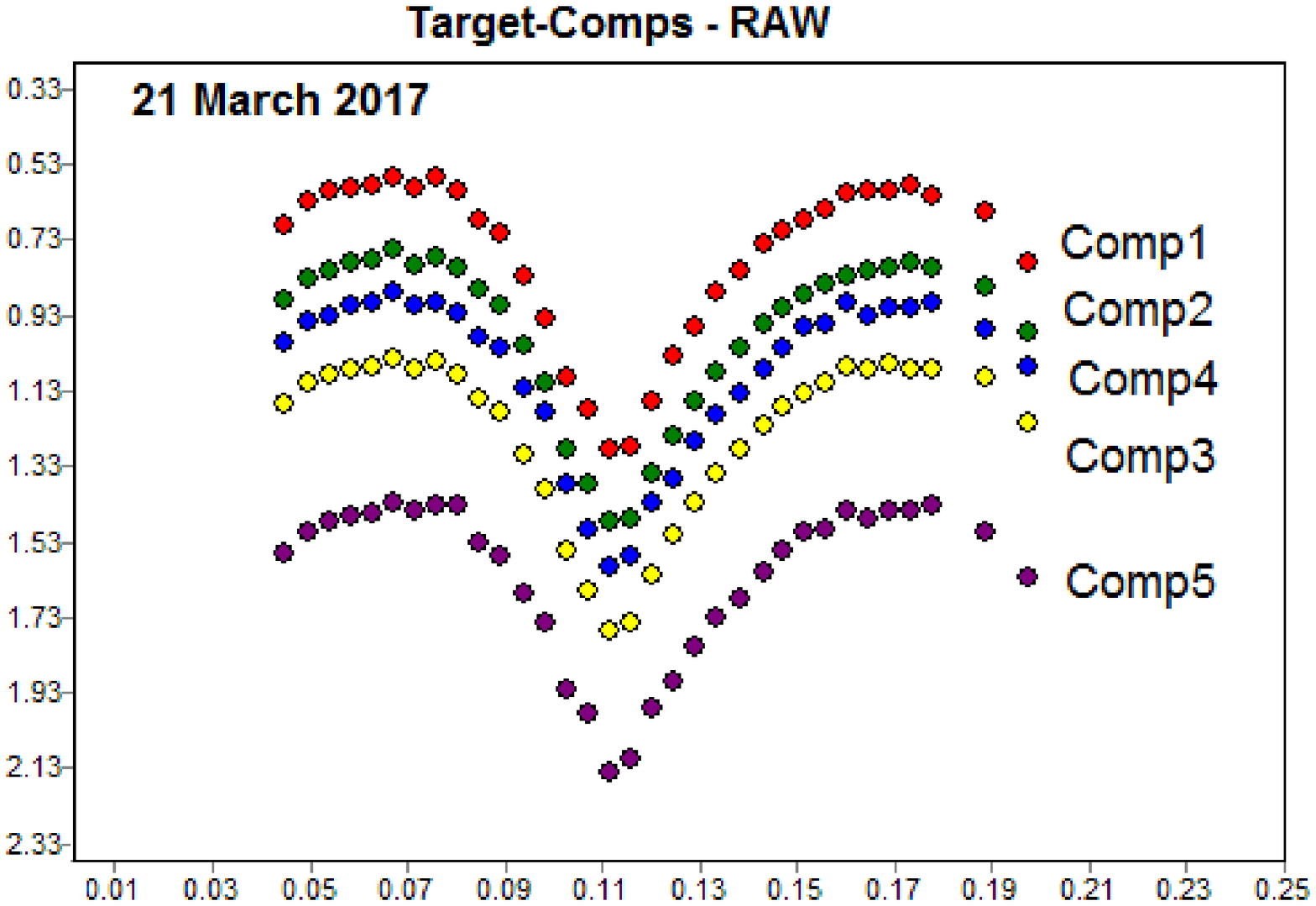, width=0.45\textwidth}}
    \caption{Derived lightcurves (plot of magnitude vs. time in days) of 9021 Fagus when using each one of the comparisons stars on 20 and 21 March 2017.}
		\label{9021March2017DC}
\end{center}
\end{figure}

The next observations of 9021 Fagus were during two neighbouring nights in 2017 March 20 and 21 on 2-m telescope. All asteroid images were taken through an R filter with exposure times between 300 and 360s. During observations the asteroid was at about 17.3 mag. The individual lightcurves from both nights are presented in Fig.\ref{9021March2017DC}. The opposition of 9021 Fagus was on February 9 and we observed  the asteroid after its opposition at the solar phase angle increased from $13.9^{\circ}$ to  $14.2^{\circ}$. 


\begin{figure}[h]
\begin{center}
	\centering{\epsfig{file=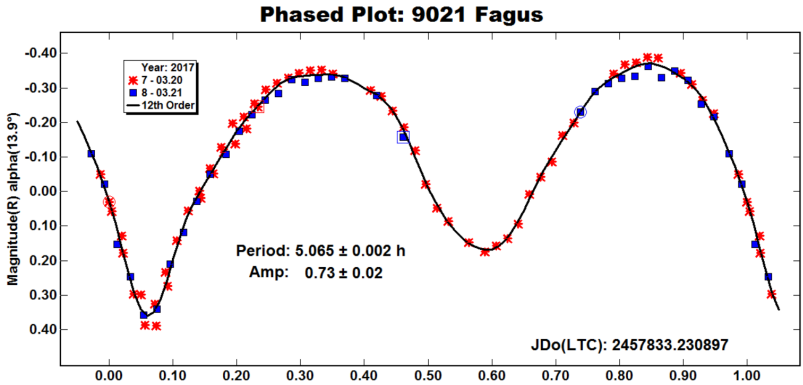, width=0.8\textwidth}}
	\caption[]{Composite lightcurve of 9021 Fagus in 2017 with Fourier fit of order 12.}
	\label{FagusCLC_001}
\end{center}
\end{figure}

Period analysis was done using MPO Canopus, which incorporates the Fourier analysis algorithm (FALC) developed by Harris et al. (1989). The estimated rotational period is $5.065 \pm 0.002$ h which has a smallest value of RMS error and best correspondence to the observational point.  Fourier fit of order 12 reveals amplitude of $0.73\pm0.02$ mag where the amplitude error is calculated as deviation from Fourier fit of all points around the highest maxima and the lowest minima. A little bit asymmetric shape of the composite lightcurve (Fig.\ref{FagusCLC_001}), with almost equal heights of the peaks but with different sharpness and different depths of the minima and high value of the amplitude, indicate an irregular shape of the asteroid.

\begin{figure}[h]
\begin{center}
	\centering{\epsfig{file=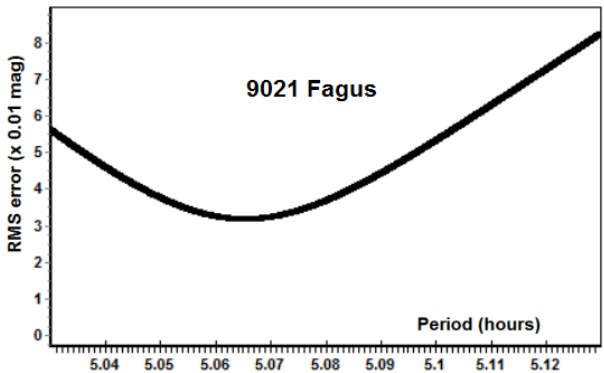, width=0.6\textwidth}}
	\caption[]{The period spectrum for 9021 Fagus based on our observations from 2017.}
	\label{spectrum}
\end{center}
\end{figure}

\begin{figure}[h]
\begin{center}
	\centering{\epsfig{file=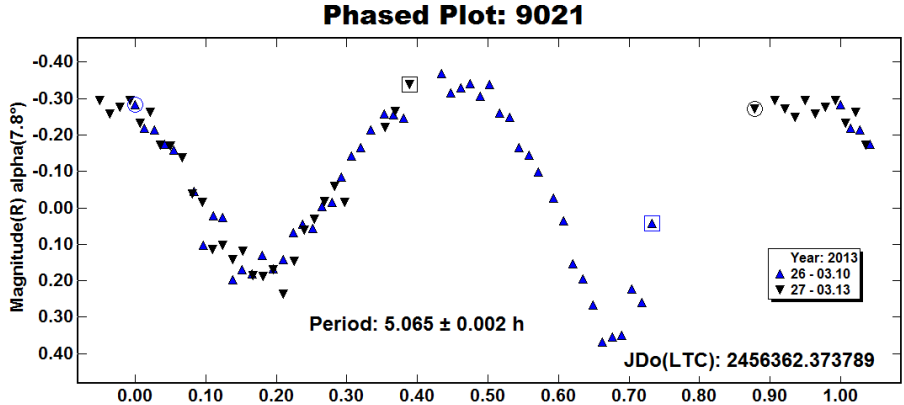, width=0.9\textwidth}}
	\caption[]{Composite lightcurve of  9021 Fagus in 2013  is  not completely covered and is constructed on the basis of the previously calculated period  of 5.065h}
	\label{FagusCLC_002}
\end{center}
\end{figure}

Using the individual (partial) lightcurves shown in Fig.\ref{9021March2013LC} and the estimated period from observations in 2017 we constructed the composite lightcurve  (Fig.\ref{FagusCLC_002}).  The amplitude of the half of the compositional lightcurve from 2013, measured using the highest maxima and the lowest  minima gives the value of $0.71 \pm 0.02$ mag.

In Table \ref{table1} the aspect data for 9021 Fagus for each night of observations are reported.  The first column is the date of the observation referring to the mid–time of the lightcurve observed. The next columns are: asteroid distance from the Sun ($r$), from the  Earth ($\Delta$), the solar phase angle, and the J2000.0 ecliptic longitude ($\lambda$) and latitude ($\beta$) of the asteroid referred to the time in the first column.

\begin{table}[h]
\begin{center}
\caption{Aspect data}
\setlength{\tabcolsep}{6pt}
\setlength{\extrarowheight}{1.5pt}
\begin{tabu}{ccccccc}
\hline
Asteroid & Date & $r$ & $\Delta$ & Phase angle & $\lambda$ & $\beta$ \\
         & (UT) & (AU) & (AU) & ($^\circ$) & ($^\circ$) & ($^\circ$) \\
\hline

9021Fagus & 2013 March 10.96 & 3.0074	& 2.0737 & 7.78	& 146.81 & -5.00 \\
				  & 2013 March 13.02 & 3.0064 & 2.0848 & 8.52 & 146.40 & -5.08 \\
					& 2017 March 20.88 & 3.0213 & 2.2528 & 13.9 & 133.50 & -2.43 \\
					& 2017 March 21.82 & 3.0211 & 2.2633 & 14.2 & 133.41 & -2.48 \\
\hline
  \end{tabu}
  \label{table1}
  \end{center}
\end{table}

\section{Conclusions}

The similar aspect data given in the Table \ref{table1} during two asteroid apparitions and almost equal values of the amplitudes of the rotational curves suggest that the rotational axis of the asteroid is in the same position relatively to the observer. The high value of the amplitude suggests that the asteroid rotation axis is almost perpendicular  to the ecliptic plane. In order to get the shape and pole positions of the asteroid it should be observed in future apparitions at various geometrical conditions (Kaasalainen et al. 2001). The rotational period of 9021 Fagus of $5.065 \pm 0.002$ h is calculated for the first time and it will give contribution to the enlargement of the database for rotational characteristics of the asteroids. We hope that our lightcurves in combination with other observations and sparse data from asteroid surveys will help  in determination of the poles and shape of the asteroid which are of significant importance for understanding the planetary system formation.

\section*{Acknowledgements}

Authors gratefully acknowledge observing grant support from the Institute of Astronomy  and Rozhen National Astronomical Observatory, Bulgarian Academy of Sciences.


\begin{thebibliography}{99}

\bibitem{Buie1998}
Buie, M., 1996, {\em ASPC, 101, 135}

\bibitem{Harris1989}
Harris, A., Young, J., Bowell, E., et al., 1989, {\em Icar, 77, 171} 

\bibitem{Ivanova2002}
Ivanova, V., Shkodrov, V., Apostolovska, G., \& Protic-Benishek, V., 2003, {\em POBeo, 75, 243}

\bibitem{Kaasalainen2001}
Kaasalainen, M., Torppa, J., \& Muinonen, K., 2001, {\em Icar, 153, 37}

\bibitem{Masiero2011}
Masiero, J., Mainzer, A., Grav, T., et al., 2011, {\em ApJ, 741, 68} 

\bibitem{Warner2009}
Warner, B., Harris, A., \& Pravec, P., 2009, {\em Icar, 202, 134, Updated 2017 April. http://www.minorplanet.info/lightcurvedatabase.html}

\bibitem{Warner2016}
Warner, B., 2016, {\em MPO Canopus software. Bdw Publishing}

\end{thebibliography}
\end{document}